\documentclass{elsart}
\usepackage{natbib}
\usepackage{amsmath,amssymb,amsfonts,euscript,changebar,subfigure,epsfig}
\usepackage{graphicx}
\usepackage{dcolumn}
\usepackage{bm}
\usepackage{epstopdf}

\begin{document}

\runauthor{Krechetnikov, Marsden and Nagib}

\begin{frontmatter}

\title{A mechanistic model of separation bubble}

\author[Pasadena]{R. Krechetnikov\thanksref{Alberta}}
\author[Pasadena]{J.~E. Marsden}
\author[Chicago]{H.~M. Nagib}

\address[Pasadena]{Control \& Dynamical Systems,
California Institute of Technology, Pasadena}
\address[Chicago]{Mechanical, Materials, and Aerospace Engineering
Department, Illinois Institute of Technology, Chicago}

\thanks[Alberta]{Current address: University of Alberta, Edmonton,
Canada}

\begin{abstract}
This work uncovers the low-dimensional nature the complex dynamics
of actuated separated flows. Namely, motivated by the problem of
model-based predictive control of separated flows, we identify the
requirements on a model-based observer and the key variables and
propose a prototype model in the case of thick airfoils as
motivated by practical applications.

The approach in this paper differs fundamentally from the logic behind
known models, which are either linear or based on POD-truncations
and are unable to reflect even the crucial bifurcation and
hysteresis inherent in separation phenomena. This new look at the
problem naturally leads to several important implications,
such as, firstly, uncovering the physical mechanisms for
hysteresis, secondly, predicting a finite amplitude instability of the
bubble, and thirdly to new issues to be studied theoretically and tested
experimentally. More importantly, by employing systematic reasoning,
the low-dimensional nature of these complex phenomena at
the coarse level is revealed.
\end{abstract}


\end{frontmatter}
\vspace{-0.5in}

\section{Introduction and methodology}

\subsection{History and motivation}

It is known that dynamic vortex shedding can lead to losses in
lift, sharp increases in drag, and destructive pitching moments
and buffeting, which all limit an aircraft flight envelope.
Therefore, in order to improve aerodynamic characteristics, flow
separation control would be highly desirable. The classical
approach---an open-loop control achieved either by mechanical or
fluidic actuation---has demonstrated robustness, but its
efficiency is still far from optimal. This standard control scheme
is based on actuator operating schedules, which are usually
constructed using extensive and costly experimental studies.

Alternatively, from a theoretical point of view, should one be
able to construct an accurate solution of the Navier-Stokes
description (NSEs) for a given airfoil shape and flow conditions,
it would suggest control strategies. However, in view of the
impossibility of solving the NSEs in real time and in view of
noisy and unpredictable real conditions, this approach is
difficult to implement. At the same time, in reality one can use
sensors on the boundary of lifting surfaces, which in turn read off
a certain amount of extra information from the physical system and
therefore should allow one to weaken the requirements on the
accuracy of theoretical prediction of the flow behavior. Thus, one
is naturally led to consider \textit{coarse} models.

However, it should be kept in mind that the dynamical behavior of
the original and coarsened (reduced) system will never be
identical, and thus one has to decide, based on the application
objectives, which aspects of the dynamics should be modelled
accurately. In this work we identify the crucial elements of the
dynamics of separation bubble, namely \textit{bifurcation} and
\textit{hysteresis}, which need to be reflected in the model and
thus result in the applicability of the model to a wide range of
physical parameters. This procedure is targeted to produce a
model, upon which an observer in a closed-loop control scheme can
be based. Being more efficient and reliable \cite{Kailath},
feedback control also naturally allows one to address the
optimization issue.

While the above is a transparent justification to appeal to coarse
modeling, the main challenge is that the resulting model should be
both \textit{low-dimensional}, for real time computational
efficiency, and \textit{physically motivated}, in order to reflect
the actual behavior for a wide range of flight and control
parameters. Since separation phenomena are clearly nonlinear, the
model should also be \textit{nonlinear}. These are certainly \textit{key
requirements} on a model.

While these key \textit{requirements} are readily appreciated, the
methods available to formulate such models are very limited and
the connection of known models to physics is rather far from what
is desired. A commonly used approach is to first generate
experimental data and, then, to extract the model by a projection
onto proper orthogonal decomposition (POD) modes using, for
example, balanced truncation or similar methods, is not reliable
in view of the open flow nature of the problem and the wide range
of governing parameters.

Because of this, we shall make use of  \textit{phenomenological}
modeling, which has been successful in many other problems, such
as the use of Duffing's equation for the buckling of elastic beams
\cite{Guckenheimer}, simple maps to describe a dripping faucet
\cite{Shaw}, which even captures the observed chaotic behavior to
a great extent, and bubble dynamics in time periodic straining
flows \cite{Kang}, to name a few. The phenomenological approach
was already used in the construction of the first few models for
separation phenomena, \textit{e.g.} \cite{Magill} and the ONERA
model \cite{Petot}, after the recent understanding of the
importance of low-dimensional models for controlling separation.
The state-of-the-art low dimensional model used in a closed-loop
control of the dynamic stall with pulsed vortex generator jets is
due to Magill \textit{et al.} \cite{Magill}. Its key feature is a
choice of the governing physical parameters, such as lift $Z$ and
separation state $B$ with $B=0$ corresponding to fully attached
flow and $B=1$ to fully separated flow. Steady states,
$B_{s}(\alpha)$ and $Z_{s}(\alpha)$, represent the baseline case
and the measured steady lift, respectively, as functions of the
angle of attack $\alpha$. The experimentally measured function
$Z_{s}(\alpha)$ which may contain a hysteretic behavior and thus
is an empirical way of accounting for a hysteresis, as suggested
by Magill \textit{et al.} \cite{Magill}. Exploiting the physical
arguments: (i) lift $Z$ $\sim$ circulation $\Gamma(\alpha)$; (ii)
relaxation to a baseline state $\lim\limits_{t \rightarrow
+\infty} B(t) = B_{s}(\alpha)$; (iii) rise in lift when a dynamic
vortex is shed $Z \sim B_{t}$, one arrives at the simplest
low-order model with adjustable parameters,
\begin{subequations}
\label{model_Magill}
\begin{align}
B_{tt} &= - k_{1} B_{t} + k_{2} \left[B_{s}(\alpha)-B\right], \\
Z_{t} &= k_{3} B_{tt} + k_{4} \left[Z_{s}(\alpha)-Z\right] +
\Gamma_{\alpha} \alpha_{t}.
\end{align}
\end{subequations}
It should be stressed that this and many earlier attempts to
develop dynamical models are based on the anzatz that this
nonlinear phenomenon behaves linearly for small variations of the
parameters involved \cite{Magill,Petot,Tobak:I}, which clearly has
many limitations, in particular cannot account for bifurcations
and hysteresis. Thus, only with an alternative approach---the
subject of this work---can one construct a model that meets the
above \textit{requirements}. As it will be clear from the text
later, while we appeal to phenomenological analysis of empirical
facts, we provide the dynamical systems grounds for it. A
symbiosis of these two methodologies yield a complete picture of
the phenomena.

\subsection{Central idea, methodology, and paper outline}

\begin{figure}[ht] \centering
\subfigure[Separated flow: open
bubble.]{\epsfig{figure=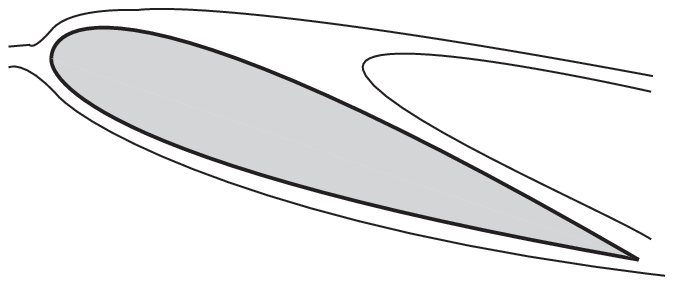,width=2.15in}\label{separated_flow}}
\qquad \subfigure[Re-attached flow: closed
bubble.]{\epsfig{figure=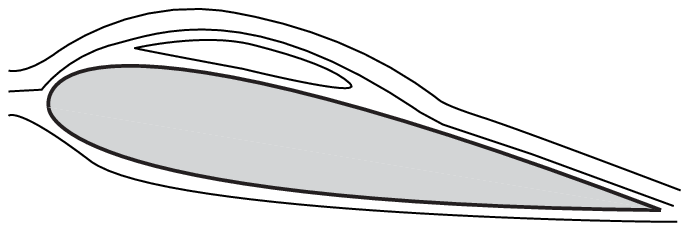,width=2.15in}\label{reattached_flow}}
\caption{On the notion of separation bubble.}
\label{notion_bubble}
\end{figure}

A central notion and object, whose dynamics we study, is \textit{a
separation bubble}, whose main features are as follows. First of
all, separation of the boundary layer develops due to an adverse
pressure gradient \cite{Schlichting} which occurs when the angle
of attack of an airfoil is sufficiently large, cf. Figure
\ref{separated_flow}, and may be followed by re-attachment as in
Figure \ref{reattached_flow}, thus forming a typical flow around
an airfoil. The region encompassed by the boundary layer is termed
a \textit{a separation bubble} after the work of Jones
\cite{Jones} and, as shown in Figure
\ref{notion_bubble}, it can be closed or open. Classification of
separation bubbles concerns their laminar or turbulent nature, but
topologically they do not differ and thus we will not be
distinguishing between various cases, but rather treat a
\textit{generic} case. It should be noted that, in certain
physical situations, a bubble needs to be understood in a
time-averaged sense \cite{Pauley}. Given the notion of a
separation bubble, our dynamical systems model will aim to capture
its characteristics, which are important for controlling
separation phenomena.

\begin{figure}[h!]
\centering
\includegraphics[scale=1.2,angle=0]{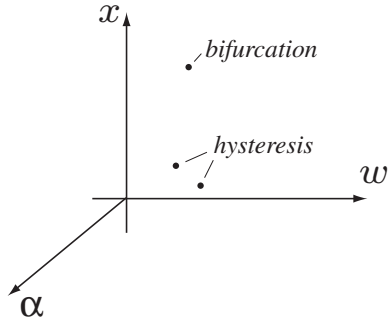}
\caption{\footnotesize A cartoon of the key dynamic elements---bifurcation and hysteresis---to be captured by the minimal number of parameters, namely the bubble
size $x$, the angle of attack $\alpha$, and the actuation amplitude $w$.}
\label{parametric_space}
\end{figure}

A central idea of this work is to approach the modeling of separation
bubble phenomena by identifying the key crucial elements of the
bubble dynamics, namely bifurcations and hysteresis, in the
appropriate portion of  parameter space, as sketched in Figure
\ref{parametric_space}. In this Figure we show the
minimal dimension of the parameter space, defined by the bubble
size $x$, the angle of attack $\alpha$, and the actuation amplitude $w$;
that is, we will be looking for the minimal model determined by
the dependence of the bubble size on the angle of attack and
actuation amplitude.

\begin{figure}[ht]
\centering
\subfigure[]{\includegraphics[scale=1.0,angle=0]{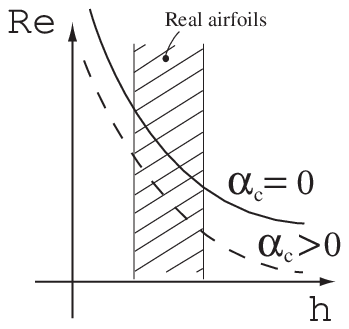}
\label{Re_vs_h}} \qquad
\subfigure[]{\includegraphics[scale=1.0,angle=0]{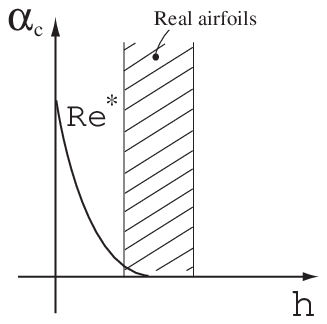}
\label{alpha_vs_h}} \qquad
\subfigure[]{\includegraphics[scale=1.0,angle=0]{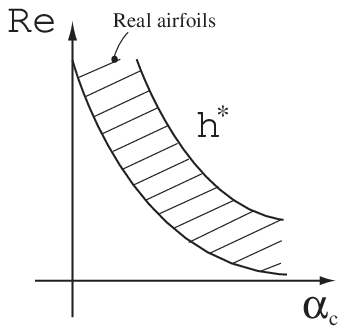}
\label{alpha_vs_Re}} \caption{\footnotesize The placement of real
airfoils in the parameter space defined by the Reynolds number
$Re$, the critical angle of attack $\alpha_{c}$, and the airfoil
thickness $h$: the critical curves corresponding to the instant
when separation occurs. $Re^{*}$ and $h^{*}$ are typical fixed
values of these parameters.} \label{thick_airfoils}
\end{figure}

This minimal approach is motivated by the fact that while
generally there are other parameters involved, such as the
Reynolds number $Re$, the critical angle of attack $\alpha_{c}$,
and the airfoil thickness $h$, the resulting model will still have
wide applicability. This can be understood based on the
aerodynamic properties of airfoils. To explain this, we draw
critical curves, i.e., when separation takes place depending upon
$Re$, $\alpha_{c}$, and $h$ in Figure \ref{thick_airfoils}.

As illustrated by Figure \ref{Re_vs_h}, in the case of real
airfoils, separation occurs at finite Reynolds numbers even at zero
critical angle of attack; the higher $\alpha_{c}$ the lower the
critical Reynolds number $Re_{c}$; also, the thicker the airfoil,
the lower $Re_{c}$. Figure \ref{alpha_vs_h} demonstrates the fact
that the thinner an airfoil, the larger the critical angle of
attack is required to achieve separation at a given Reynolds
number $Re^{*}$. Finally, in the $\alpha_{c}$-$Re$ plane in Figure
\ref{alpha_vs_Re} one can observe that for fixed airfoil thickness
$h^{*}$ separation can occur at zero $\alpha_{c}$, which requires
high enough Reynolds numbers. Since in reality the Reynolds
numbers are huge (e.g. for real aircraft $Re$ varies between $10^{6}$ and $10^{11}$), one concludes that limiting ourselves to ``thick airfoils'', which
can, in fact, be regarded as real airfoils since they have to carry structural
load and fuel, is not a serious restriction in this first step
towards low-dimensional modeling of separation phenomena.

To achieve the above objectives of our modeling identified above, we
will appeal to the tools of the bifurcation and catastrophe theory
\cite{Arnold:I}, as will be made precise in \S
\ref{sec:bifurcation_bubble}. The outline of the paper is as
follows. In \S \ref{sec:bifurcation_bubble}, we discuss the first
nonlinear aspect of separation bubbles, namely bifurcation
phenomenon and the way to model it. In \S
\ref{sec:hysteresis_bubble}, we explore the basic physics of
hysteresis phenomena, and suggest its mathematical model and how
to construct a single model capable of capturing both bifurcation
and hysteresis.

\section{\label{sec:bifurcation_bubble}Bifurcation in the dynamics of separation bubble}

\subsection{\label{subsec:notion_bifurcation}On the notion of bifurcation}
\begin{figure}[h!] \centering
\subfigure[Open bubble, $x_{r} = \infty$: under-actuated case,
$w<w_{c}$
.]{\epsfig{figure=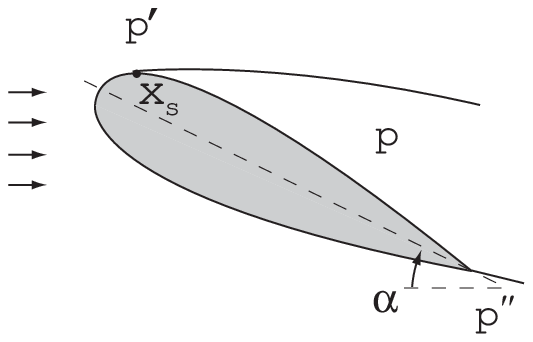,width=1.95in}\label{open
bubble}} \qquad \subfigure[Closed bubble, $x_{r} < \infty$:
controlled case, $w>w_{c}(\alpha)$
.]{\epsfig{figure=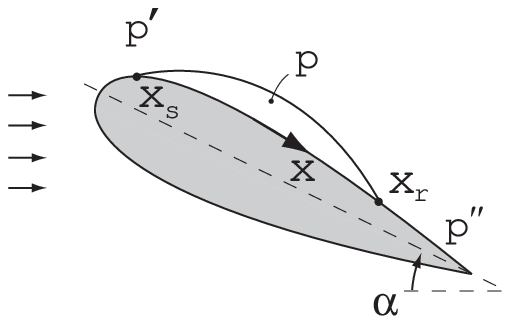,width=1.95in}\label{closed_bubble}}
\caption{Basic setup and primary bifurcation.}
\label{basic_bifurcation}
\end{figure}

As was noted in the introduction, bubbles can
be either in a closed or open state. This allows us to introduce a
key element of the low-dimensional modeling, namely it must
capture this \textit{basic bifurcation} from an \textit{open} to a
\textit{closed} state, as shown schematically in Figure
\ref{basic_bifurcation}, which is also known as \textit{bursting}
\cite{Tani}.

Notably, the fact that this is the primary bifurcation was
realized just recently \cite{Ghil}. The vast literature on
separation bubbles behavior is still at a descriptive level and
suggests that one separated flow is not like any other. Here we
take a different point of view, i.e. we treat the coarse behavior
of separation bubbles as (generic) phenomena that can be modelled
by a single low dimensional dynamical system.

\subsection{\label{subsec:quantification}Quantifying separation bubbles}

To quantify the behavior of a separation bubble, consider the
coordinate $x$, measuring the distance along the airfoil from the bubble onset to the bubble reattachment, as shown
in Figure \ref{basic_bifurcation}. The bubble dynamics in the
first approximation can be described by two parameters: the location
of separation, $x_{s}$, and of reattachment, $x_{r}$, which
can move under the change of flight and control parameters. In
some cases, e.g. the Glauert Glas II airfoils, the separation
point $x_{s}$ remains fixed for all practical purposes. Therefore,
we will start by considering only the behavior of the reattachment
point, which experiences a primary bifurcation in the above sense;
extending the model to include variation of $x_{s}$ will require
the addition of a reliable separation criterion. As an alternative
to $x_{r}$, one could also utilize the bubble area. From now on we
will use $x$ as a variable representing the bubble state.

\subsection{\label{subsec:bubble_bifurcation_nature}On the physical nature of bifurcation}

\begin{figure}[ht] \centering
\subfigure[No
excitation.]{\epsfig{figure=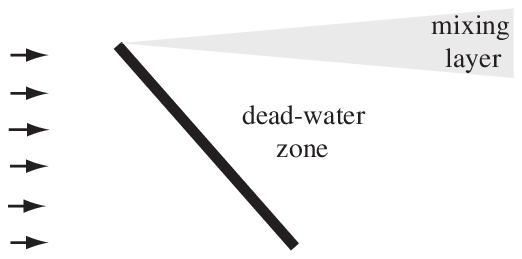,width=1.65in}\label{flap_I}}
\quad \subfigure[Weak
excitation.]{\epsfig{figure=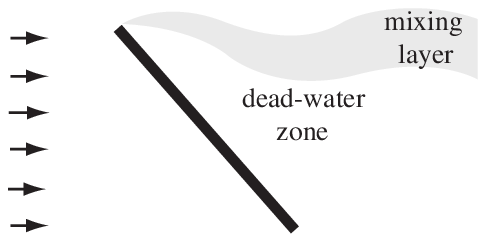,width=1.65in}\label{flap_III}}
\quad \subfigure[Strong
excitation.]{\epsfig{figure=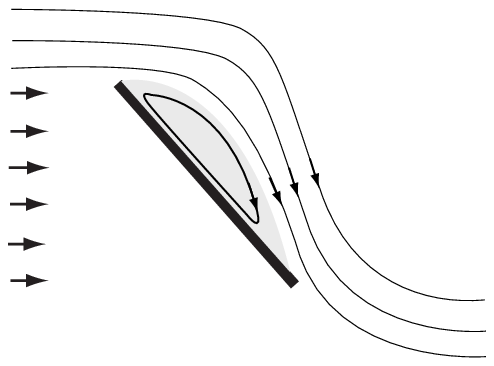,width=1.65in}\label{flap_II}}
\caption{On the mechanism of actuation.}
\label{mechanism_actuation}
\end{figure}

The mechanism by which the excitation affects the flow lies in the
generation of instabilities, and thus of Large Coherent Structures
transferring high momentum fluid towards the surface, and
therefore leading to reattachment, as indicated in Figure
\ref{mechanism_actuation}. Since actuation exploits the
instabilities of the shear layer \cite{Oster}, the response to
actuation depends on both $w$ and $\omega$ and therefore is
nonlinear.
The latter again indicates, now from the point of view of
actuation control mechanisms, that the low-dimensional model must
be nonlinear.

\begin{figure}[h!] \centering
\epsfig{figure=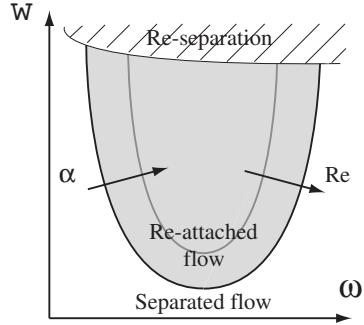,height=1.75in} \caption{Effect of
time-variant actuation: criticality of actuation amplitude $w$ and
frequency $\omega$. Shaded region corresponds to reattached flow
(closed bubble). Arrows indicate the change in location of the
transition curve with increasing $Re$ and
$\alpha$.}\label{criticality}
\end{figure}
As follows from experiments, the critical phenomena are as
sketched in Figure \ref{criticality}, where the shaded region
corresponds to a reattached flow (that is, a closed bubble) and the arrow
indicates a change in location of the transition curve with an
increase in $\alpha$. The size of the bubble, $x$, has a specific
dependence on the amplitude $w$ and frequency $\omega$ of
actuation, \textit{i.e.} ${\partial x /
\partial w} < 0, \ {\partial x /
\partial \omega} < 0$, when moving away from the origin $(w,\omega)=\mathbf{0}$ in Figure
\ref{criticality}. In this work we focus on the case of
time-invariant actuation, $\omega=0$, although the time-varying case
will be commented on later in this section.

\begin{figure}[ht] \centering
\subfigure[Hump
model.]{\epsfig{figure=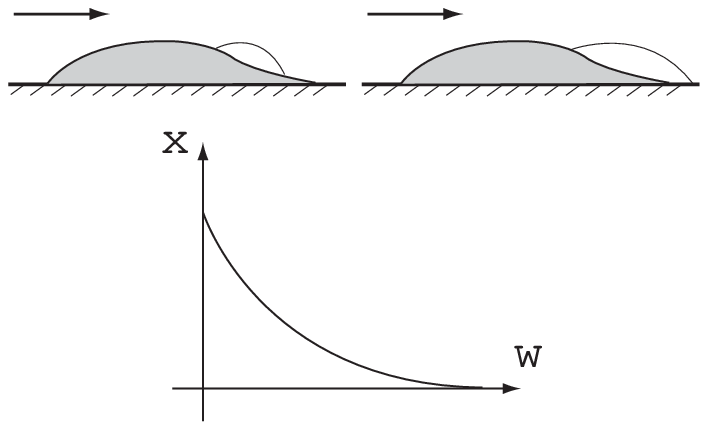,width=2.25in}\label{hump_model}}
\qquad \subfigure[Airfoil
model.]{\epsfig{figure=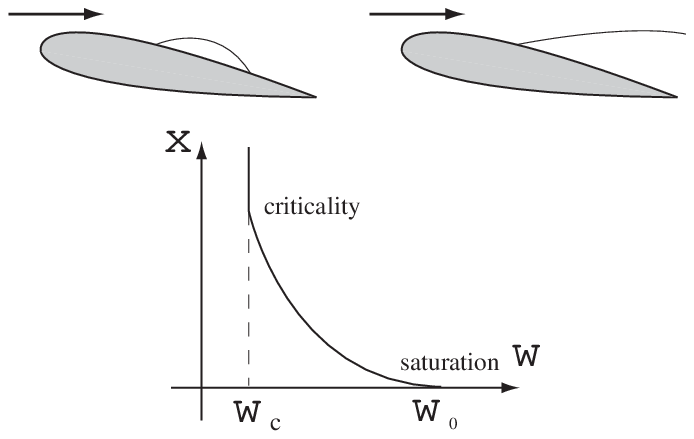,width=2.25in}\label{airfoil_model}}
\caption{Two basic experimental configurations.}
\label{experimental_configurations}
\end{figure}

Finally, it is notable that the criticality and hysteresis
phenomena depend on the connectedness of the flow domain: the
bubble experiences bifurcation only in the case of flow around an
airfoil, as in Figure \ref{airfoil_model}, while in the case of a
hump model in Figure \ref{hump_model}, which is frequently used in
experiments, there is no bifurcation. Thus, there are two basic
configurations in which the behavior of the separation bubble
differs: the \textit{hump model}  and the \textit{airfoil model}.
Namely, in the hump case $x(w)$ is smooth, while in the case of an
airfoil $x(w)$ is discontinuous. Also, as will be important in \S
3, the hysteresis phenomena are present only in the airfoil case.
Here, in view of its practical importance, we naturally focus on
the airfoil case.

\subsection{\label{subsec:bubble_bifurcation_model}Modeling the bubble bifurcation}

In developing a model, we are guided by the principle of a minimal
complexity together with the physical requirements one has to
meet. At the methodological level, there are two basic ways to
account for the form of $x(w)$, which has both the saturation and
criticality shown in Figure \ref{experimental_configurations}: (a)
to design $F(x,w)=0$ as an \textit{algebraic} relation, or (b) to
introduce a \textit{dynamic} description
$F(x,\dot{x},\ddot{x},\ldots,w)$. The latter approach is better
suited for dynamics and control purposes, because in the case of
active feedback control one would need to deal with a few
characteristic times and transient effects, and thus the model
should be time-dependent. The simplest possible way of introducing
time-dependent dynamics is a second-order oscillator model,
$\ddot{x} - \mu \dot{x} = F(x,w)$, where $\mu$ is a damping
parameter. The justification for the latter may serve the fact
that both separation and reattachment points may oscillate
\cite{Pauley}.

In what follows, we first formulate mathematical requirements on a
model in \S \ref{subsubsection:math_req}, then by appealing to the
ideas of a potential function in \S \ref{subsubsection:pot_fun}
and a dynamic bifurcation in \S \ref{subsubsection:dyn_bif}, we
construct the model in \S \ref{subsubsection:model}.

\subsubsection{\label{subsubsection:math_req}Mathematical requirements}

\begin{figure}
\centering
\subfigure[Uncontrolled]{\epsfig{figure=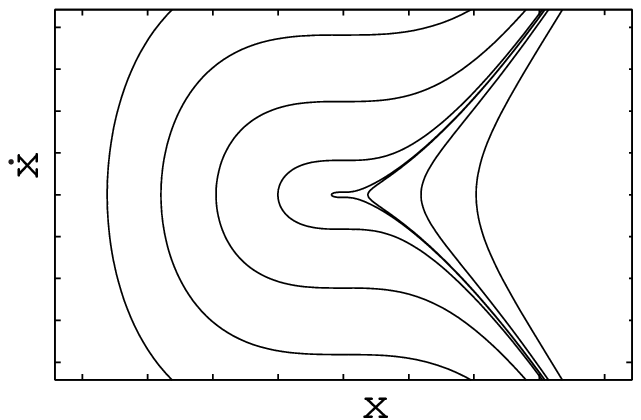,width=1.75in}\label{open}}
\quad
\subfigure[Controlled]{\epsfig{figure=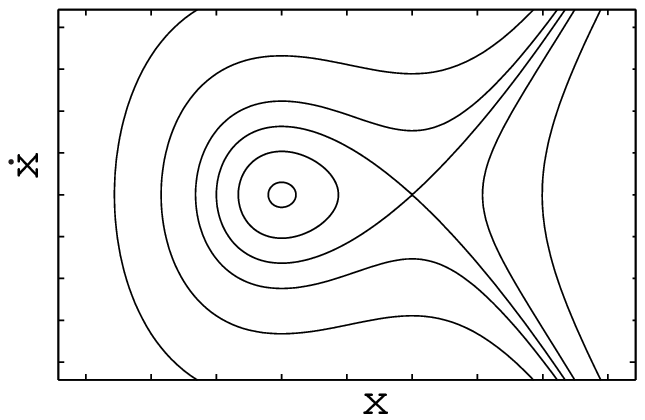,width=1.75in}\label{closed}}
\caption{Takens-Bogdanov
bifurcation.\label{Takens_Bogdanov_bifurcation}}
\end{figure}

Naturally, the bubble size $x$ also depends on a \textit{flight
parameter}, in our case the angle of attack $\alpha$, which needs
to be incorporated in the model; thus, $F=F(x,w,\alpha)$. Since we
want to minimize the functional complexity, but to retain the
nonlinear features of the phenomena, the simplest form is a
quadratic nonlinearity, $F(x,w,\alpha) = x^{2} + b(w,\alpha) \, x
+ c(w,\alpha)$, which possesses a Takens-Bogdanov bifurcation, as
shown in Figure \ref{Takens_Bogdanov_bifurcation}, when $b^{2} - 4
c$ changes sign.

Indeed, equilibria points are given by $x_{1,2} = - \frac{b }{2} \pm \frac{1 }{ 2} \sqrt{b^{2} - 4 c}$, so that $F$ can be
represented as $(x - x_{1}) (x - x_{2})$. The eigenvalues of the
linearization around $x=x_{1}$ are given by $\lambda^{2}_{1,2} =
x_{1}-x_{2} = \sqrt{b^{2}-4 c}$, while the eigenvalues of the
linearization around $x=x_{2}$ are $\lambda^{2}_{1,2} =
x_{2}-x_{1} = - \sqrt{b^{2}-4 c}$. Thus when $b^{2}-4 c$ changes
sign, one observes the transition from the picture in Figure
\ref{open} to the one in Figure \ref{closed}. The requirements on
the parameters in this model are dictated by the physics:
\begin{description}
\item[(a)] The stability of equilibria points should obey
\begin{align}
&\alpha < \alpha_{c}: \ & &b^{2} - 4 c > 0 \ \mathrm{(stability)}, \nonumber \\
&\alpha > \alpha_{c}: \ & &w > w_{c}(\alpha), \ b^{2} - 4 c > 0 \ \mathrm{(stability: \ no \ separation)}, \nonumber \\
& & &w<w_{c}(\alpha), \ b^{2} - 4 c < 0 \ \mathrm{(instability: \
separation)}, \nonumber
\end{align}
where stability implies that one equilibrium is stable ($\lambda$
is imaginary), and the second one is unstable ($\lambda$ is real).
The above inequalities indicate that the physical behavior of the
model is also governed by the critical angle of attack
$\alpha_{c}$, when the flow separates at $w=0$, and the critical
control amplitude $w_{c}(\alpha)$, when flow reattaches at
$\alpha$ fixed.

\item[(b)] The critical actuation amplitude $w_{c}$ should grow with $(\alpha-\alpha_{c})$,
since the higher the angle of attack, the more control input is
required to make the flow reattached.

\item[(c)] The bubble size $x$, which is a stable equilibrium, should shrink, $x \rightarrow 0$ as $w$ increases. At the same time, the domain of attraction of
this equilibrium should shrink too, so that the bubble becomes
susceptible to the finite-amplitude instabilities, as it is known
from experiments, cf. the upper part of Figure \ref{criticality}.
\end{description}

\subsubsection{\label{subsubsection:pot_fun}Potential function approach}

\begin{figure}[ht]
\centering \subfigure[Potential function for a finite bubble.]
{\includegraphics[scale=1.1,angle=0]{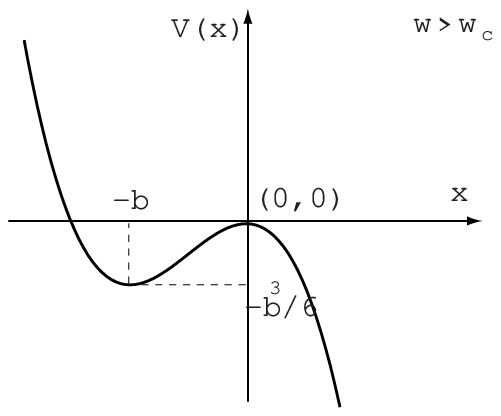}\label{potential_stable_bifurcation}}
\qquad \subfigure[Potential function for an infinite bubble.]
{\includegraphics[scale=1.1,angle=0]{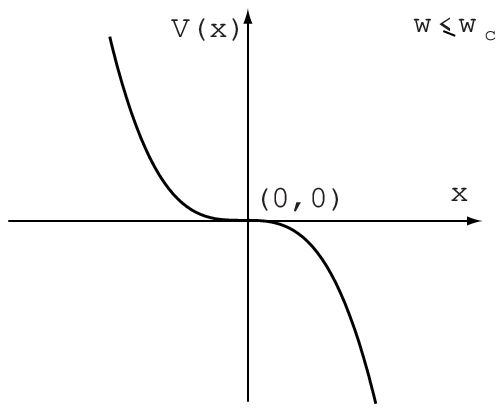}\label{potential_unstable_bifurcation}}
\caption{\footnotesize Potential function $V(x)$; $d=0$, $c=0$ in
\eqref{potential_bifurcation_formula}.\label{potential_bifurcation}}
\end{figure}

In order to get better insight in the model construction, let us
assign a potential function $V(x)$ such that $V^{\prime}(x) = -
F(x)$:
\begin{align}
\label{potential_bifurcation_formula} V(x)= \frac{x^{3} }{ 3} + b(w)
\frac{x^{2} }{ 2} + c(w) x + d(w),
\end{align}
which is physically determined by the elastic properties of a
bubble and its interaction with the outer flow. Then we can
observe that a finite bubble corresponds to $V(x)$ as in Figure
\ref{potential_stable_bifurcation}, and an infinite bubble
corresponds to Figure \ref{potential_unstable_bifurcation}.

\begin{figure}[h!] \centering
\includegraphics[scale=1.45,angle=0]{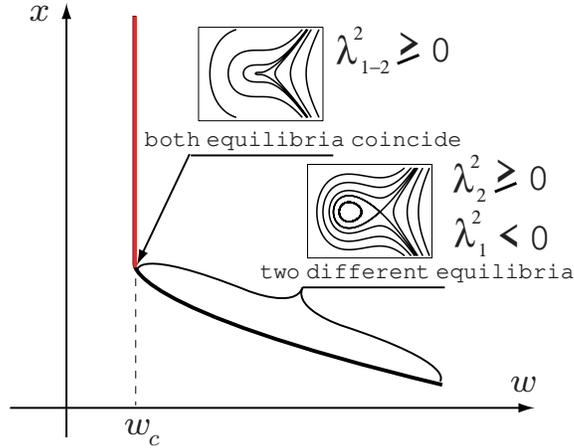}
\caption{\footnotesize Critical curve in the $(x,w)$-plane: on the
dynamic bifurcation; solid black line represents stable
equilibria, solid red line is a dynamic bifurcation when bubble
grows indefinitely with time. Phase portraits in rectangles
correspond to the ones in Figure
\ref{Takens_Bogdanov_bifurcation}.} \label{bifurcation}
\end{figure}

Without loss of generality, we can assume that $d = 0$. Considering $w$, as a control parameter,
the requirements on the coefficients in $V(x)$ are such that the
equilibria, $V'(x)=-x^{2}-b x - c=0$, obey
\begin{align}
&w > w_{c}: \ \mathrm{two \ equilibria \ (stable \ and \
unstable)}, \ V''(x_{1}) > 0, \
V''(x_{2}) < 0; \nonumber \\
&w \le w_{c}: \ \mathrm{only \ one \ equilibrium \ point, \ which
\ is \ unstable \ (marginally \ stable)}. \nonumber
\end{align}
The stability conditions can also be reformulated in terms of
eigenvalues, as indicated in Figure \ref{bifurcation}. In this
particular case, the equilibria points $x_{i}(w,w_{c})$ are easily
computable: $x_{1}=-b$ and $x_{2}=0$. The stability criterion
(second variation) for these equilibria is given by the sign of
the second derivative, $V''(x)= - 2 x - b$, which at the
equilibria points assumes the values $-b$ and $b$, respectively.
Besides the stability conditions, one needs to impose ${\mathrm{d}
x_{1} / \mathrm{d} w} < 0$, since the bubble shrinks when the
control amplitude increases. Thus, the bifurcation from the state
in Figure \ref{potential_stable_bifurcation} to the one in Figure
\ref{potential_unstable_bifurcation} is obviously associated with
the condition when $b(w_{c})=0$. As one can further infer, in the
space of curves in $(w, w_{c})$ there is an infinite number of
solutions $b=b(w,w_{c})$, $c=c(w,w_{c})$. In practice, a
\textit{systematic procedure} would be as follows: depending on
the particularities of the experimental data, one expands $b$ and
$c$ in terms of some basis functions of $w$, $w_{c}$, etc., and
then determines coefficients in that expansion through the
calibration procedure.

\subsubsection{\label{subsubsection:dyn_bif}Dynamic bifurcation}

The transition from one potential to another is controlled by a
bifurcation parameter, such as angle of attack $\alpha$ or
actuation amplitude $w$. In fact, the latter two parameters are
interchangeable to a certain extent as argued in \cite{Amitay},
since a change in $\alpha$ or in $w$ leads to a change in
circulation around an airfoil, and thus to a change in the flow
structure. Apparently, this transition of $x$ from finite to
infinite is \textit{dynamic} in a sense that the bubble becomes
infinite in Figure \ref{potential_unstable_bifurcation} as time $t
\rightarrow \infty$. This \textit{dynamic bifurcation} can be
clarified using phase portraits in Figure \ref{bifurcation}, and
should be opposed to the standard static bifurcation, which is of
algebraic nature as resulting from the condition of vanishing
vector fields. As one can learn from Figure \ref{bifurcation}, at
the critical value of the actuation amplitude $w_{c}$ both
equilibria coincide and are unstable (marginally stable), so that
the bubble grows with time and becomes unbounded for $t
\rightarrow \infty$. For $w > w_{c}$ there are two equilibria
points, one is unstable and one is stable. The latter corresponds
to the situation when bubble is of finite size, and this state has
a finite domain of attraction. Note that the potential energy
shape, as in Figure \ref{potential_bifurcation}, is crucial in
allowing the ``dynamic'' bifurcation: a $V$-shaped potential
function apparently would not allow this type of bifurcation, as
well as the domain of attraction would be modelled inconsistently
with physics. Similar type of argument will be applied to the
hysteresis phenomena in \S \ref{sec:hysteresis_bubble}. In
conclusion, having identified, based on the physical
argumentation, that the potential should be of the shape as in
Figure \ref{potential_bifurcation} in order to allow a dynamic
bifurcation, the problem reduces to determination of the
coefficients in \eqref{potential_bifurcation_formula}. This
general procedure is the subject of the catastrophe theory
\cite{Arnold:I} and, at the technical level, is in the realm of
calculus \cite{Marsden:I}.

\subsubsection{\label{subsubsection:model}Model and its interpretation}

For simplicity, restricting ourselves to the case of thick
airfoils when separation occurs at $\alpha_{c}=0$ without
actuation, with one of infinitely many admissible choices of $b$
and $c$ we get:
\begin{align}
\ddot{x}= -V_{x}(x;\alpha,w) = -\mu \dot{x} + (x-\alpha)^{2} +
f(w) \, x. \label{simplest_model}
\end{align}
Here $f(w) = a_{1} w + a_{2} w^{2} + \ldots$ represents a
\textit{nonlinear response} of the separation region to actuator
excitations $w$. The product $f(w) \, x$ implies that the effect
of actuation depends upon the bubble size $x$. As required,
$f(w_{c}) = 2 \, \alpha^{1/2}$ and the bubble shrinks as $\sim
f^{-1}$ for $w \rightarrow \infty$. While this is the simplest
possibility, from the above description it is clear that there is
enough flexibility to calibrate the model through the fitting
functions, $F(x,w,\alpha)$, and parameters, $(b,c,\ldots)$ within
the given above bounds.

By construction, the model \eqref{simplest_model} reflects the basic
generic dynamic behavior of separation bubbles. In the
\textit{conservative} time-invariant case the parameter space is
just $(\alpha,f(w))$. When control is absent, $f(w)=0$, the bubble
is open, which corresponds to an unstable phase portrait in Figure
\ref{open}, that is any initial conditions lead to an unbounded
bubble size $x$. When sufficient control is applied (consider first
$\alpha$ fixed), the bubble closes, which is reflected in the
change of the phase portrait as shown in Figure \ref{closed}. In
this case there are two equilibrium points, one is a saddle, which
is unstable and thus not physically observable, and another one is
a stable center. Therefore, there exists a non-zero basin of
attraction which leads to a finite bubble size, $x < \infty$.
Figure \ref{closed} also suggests that the system is susceptible
to finite-amplitude instability for $w > w_{c}$, the fact which is
conceivable but has never been studied in experiments
systematically. Nevertheless, it is known empirically that the bubble
opens if the actuation amplitude becomes large enough, as in
Figure \ref{criticality}; see also \cite{Krechetnikov}. Also, the
fact that the boundary layer is susceptible to finite-amplitude
instabilities \cite{Nayfeh} suggests that the separation bubble
formed out of it may also be finite-amplitude unstable. The
inclusion of dissipation in the model \eqref{simplest_model} does not
change the nature of the phase portrait; however, it does change
the basin of attraction.

Finally, the inclusion of time-varying effects in the control, $w
= w_{0} \cos{\omega t}$ with $\omega \neq 0$, also demonstrates
that the bubble transforms from an open to a closed state. Thus,
as required, the model \eqref{simplest_model} captures the primary
bifurcation and dynamic behavior of the separation bubble, except
for the hysteresis. In the rest of this paper we will explain how
the model \eqref{simplest_model} can be {\it enhanced} to account
for the hysteresis shown in Figure \ref{conjecture}. While the
model \eqref{simplest_model} is given for one of infinitely many
choices of parameters, it is clear that a variety of other
admissible choices can produce the same type of bifurcation and
dynamics. This freedom to choose parameters is important, however,
in order to fit the model to a particular application via
calibration.

\subsection{\label{subsec:analogy_bifurcation}Analogy to other physical phenomena}

\begin{figure}[h!] \centering
\subfigure[Four-roll mill
\cite{Taylor:II}.]{\epsfig{figure=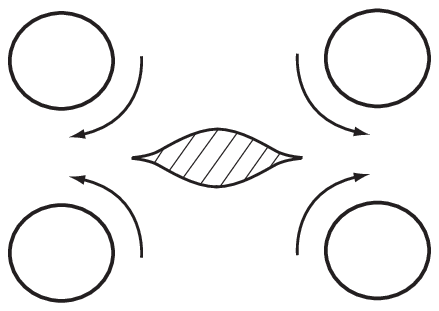,width=1.45in}\label{fourroll}}
\qquad \subfigure[Codim-2
bifurcation.]{\epsfig{figure=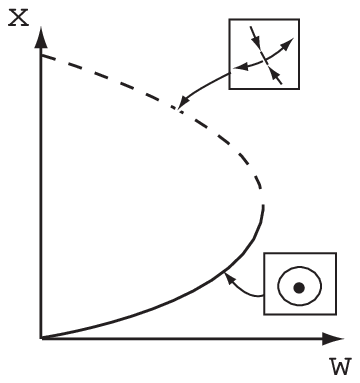,width=1.75in,height=1.75in}\label{codim2_bifurcation}}
\caption{Bubble deformation in a four-roll mill \cite{Kang}.}
\end{figure}

It is notable that a model of a similar form was deduced
\textit{ad hoc} for a real bubble deforming in a straining flow
studied by Kang \& Leal \cite{Kang}, shown in Figure
\ref{fourroll}, which experiences a bifurcation from a deformed
but closed state to an open tip-streaming state, when bubble forms
pointed open ends emitting tiny bubbles.

Namely, the model is $\ddot{x}= -\mu \dot{x} + (a x - b x^{2}) +
w$, where $w$ is the control parameter (Weber number). The
dynamics of this problem is illustrated in Figure
\ref{codim2_bifurcation} for the conservative case, $\mu=0$, and
reflects the fact that for the same $w$ there are two equilibria,
one of which is a stable center and another one is a saddle; the
latter is not observed physically in view of its unstable
character. This problem also illustrates the analogy of the
dynamics of real and separation bubbles.

\section{\label{sec:hysteresis_bubble}Hysteresis in the dynamics of separation
bubble}

\subsection{\label{subsec:experimental_observations_hysteresis} Experimental
observations: the model objectives}

\begin{figure}[h!] \centering
\includegraphics[scale=1.0,angle=0]{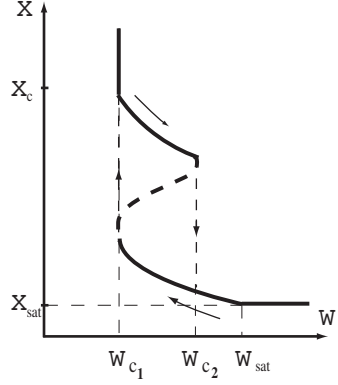}
\caption{\footnotesize Experimental effect of time-varying actuation: hysteresis phenomena in
amplitude $w$.}
\label{conjecture}
\end{figure}

The basic effects of time-varying control were discussed in \S
\ref{subsec:bubble_bifurcation_nature} and reflected in Figure
\ref{criticality}. However, the effect of changing amplitude and
frequency is not trivial in view of the presence of a hysteresis
\cite{Nishri,Greenblatt:I} in all variables $(\alpha,w,\omega)$,
as illustrated in Figure \ref{conjecture} for the dependence of
the bubble size on the actuation amplitude, $x(w)$. Experiments
demonstrate that hysteresis is present no matter how slowly the
actuation amplitude $w$ is changed, which suggests that the model
should depend only on the sign of the rate $\dot{w}$.

Therefore, the model should reflect the fact that there are two
stable steady state solutions for the range of the control
parameter $w_{c_{1}} < w < w_{c_{2}}$, as in Figure
\ref{conjecture}, which is an experimental fact. Mathematically,
this means that the selection between these two solutions is due
to the placement of initial conditions in the corresponding domain
of attraction. Also, for $w > w_{c_{2}}$, there should be only on
stable solution, while for $w < w_{c_{1}}$ the bubble should
``bifurcate'' to infinity in a dynamical manner as described in \S
\ref{subsec:bubble_bifurcation_model}. The challenge of modeling
the hysteresis comes from the fact that the behavior of the bubble
is known only from experimental observations, while there are no
analytical results. Clearly, the domain of attraction of stable
solutions is not well-characterized from the existing empirical
data.

Therefore, in order to model hysteresis, one first needs to
understand its physical origin, which is addressed below, in \S
\ref{subsec:hysteresis_physics}, where we suggest the physical
mechanisms of the hysteresis. This together with the dynamical
systems and catastrophe theory allow us to modify the model
\eqref{simplest_model} to account for hysteresis, which is the
subject of \S \ref{subsec:hysteresis_model}.

\subsection{\label{subsec:hysteresis_physics}On the physics of hysteresis}

Physically, the separation bubble is caused by a strong adverse
pressure gradient, which makes the boundary layer separate from
the curved airfoil surface. Actuation with $w>w_{c}$ effectively
reduces the adverse pressure gradient \footnote{Note that for some
airfoils the same effect can be achieved by changing the angle of
attack $\alpha$, i.e., the larger the angle of attack $\alpha$,
the stronger the adverse pressure gradient: this
``interchangeability'' of the effects of the actuation amplitude
and the angle of attack is well-known \cite{Amitay} and is
reflected in the dependence $w_{c}(\alpha)$.} and makes the bubble
closed, as in Figure \ref{closed_bubble}.  This can be seen from
Bernoulli's equation, since the velocity drop is related to the
pressure rise, $p + {\rho u^{2} / 2} = p_{0}$, where $p$ is a
dynamic pressure, and $p_{0}$ is the fluid pressure at rest. From
Bernoulli's equation and Figure \ref{basic_bifurcation} we can
conclude that pressure rise $p_{i}^{\prime}-p_{i}^{\prime\prime}$
and bubble length $l$ correlate
$p_{1}^{\prime}-p_{1}^{\prime\prime} <
p_{2}^{\prime}-p_{2}^{\prime\prime}$ and $l_{1} < l_{2}$,
respectively. For the current purposes we neglect by the second
order effects of vorticity and thus assume constant pressure
inside the bubble, $p_{1}=p_{2}=\mathrm{const}$.
\begin{figure}[h!]
\centering \subfigure[Separation
bubble.]{\includegraphics[scale=1.0,angle=0]{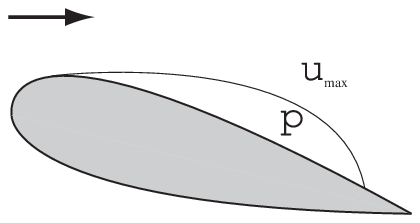}\label{hydrofoil}}
\qquad \subfigure[Static
bubble.]{\includegraphics[scale=1.0,angle=0]{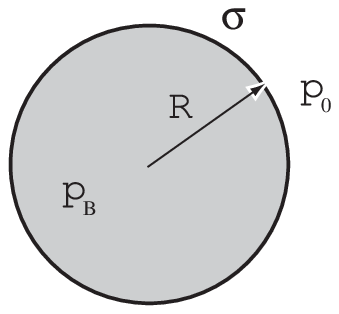}\label{static_bubble}}
\caption{\footnotesize Contrasting separation and ordinary
bubbles.} \label{contrasting}
\end{figure}

Let us compare the above behavior of a separation bubble with that
of a real static bubble (see Figure \ref{contrasting}), which is
governed by $p_{B} = {2 \sigma / R} + p_{0}$, where $p_{B}$ is the
pressure inside the bubble, $p_{0}$ is the pressure outside the
bubble, $\sigma>0$ is the interfacial tension, and $R$ is the
radius of the bubble. Apparently, if the pressure $p_{0}$ outside
the bubble decreases while the pressure inside is maintained
constant, the bubble shrinks \footnote{In reality, the pressure
inside in inversely proportional to the bubble radius  which
results in bubble growth.}. This behavior of a real bubble, when
its pressure inside is maintained constant, contrasts with that of
a separation bubble, which grows if the pressure outside the
bubble, $p$, reduces. The underlying physics of these two problems
differs: in the first case, the phenomena are dominated by static
forces, while separation phenomena are dynamic.
\begin{figure}[h!] \centering
\subfigure[Closed
bubble.]{\epsfig{figure=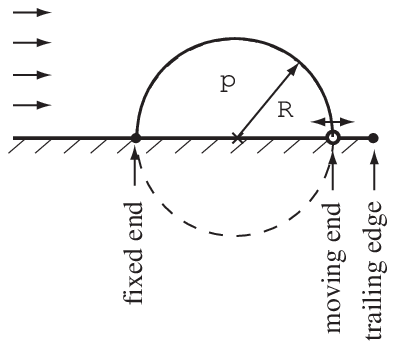,width=1.75in}\label{model_a}}
\qquad \subfigure[Open
bubble.]{\epsfig{figure=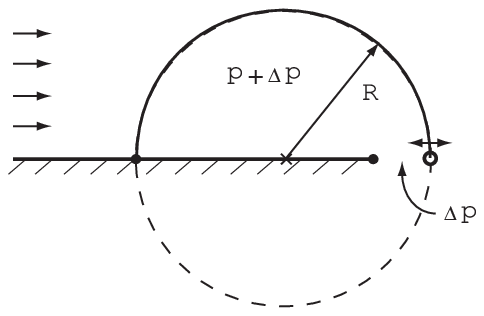,width=1.75in}\label{model_b}}
\caption{Mechanical model of separation bubble hysteresis.}
\label{mechanical_analog}
\end{figure}

This suggests that the separation bubble boundary possesses
elastic properties, which for the current purposes can be modelled
with negative interfacial tension. Note that real (positive)
tension tends to minimize the interfacial area, while the
effective tension of the shear layer tends to maximize the bubble
boundary and only the external energy input (excitation)
counterparts this effect and makes the bubble closed: this
justifies a negative sign of the tension. Alternatively, one can
use a nontrivial state equation for the pressure inside the
bubble, which can be measured experimentally. The elasticity of
the separation bubble is evidenced by introducing disturbances
outside the bubble and observing the changes in the bubble
characteristics, i.e. shape and pressure inside\footnote{Both
elasticity and non-trivial state equation of the separation bubble
have been confirmed experimentally (personal communication: John
Kiedaisch).}.

With the above physical background, we can provide a simple
mechanistic model explaining the origin of the hysteresis. For
simplicity, consider the two-dimensional situation depicted in
Figure \ref{mechanical_analog}: a hemispherical bubble having
variable size with the left end fixed and with its right end free
to move thus modeling a separation bubble with moving reattachment
point. The bubble size changes depending upon the free-stream
velocity $u_{\mathrm{max}}$, which is chosen to be the control
parameter. When $u_{\mathrm{max}}$ increases and the right end of
the bubble reaches the trailing edge at $R_{0}$ at critical
$u_{\mathrm{cr}}^{2}$, the pressure inside the bubble increases by
a finite amount, $p_{0} \rightarrow p_{0} + \Delta p_{0}$, which
is due to suction of a high pressure fluid from the lower side of
the airfoil. Hence the bubble size increases abruptly by some
amount. Conversely, when $u_{\mathrm{max}}$ decreases and bubble
reaches the trailing edge at $R_{0}$ at a different critical
$u_{\mathrm{cr}}^{1}$, the pressure inside the bubble relaxes to
its original value, $p_{0} \rightarrow p_{0} - \Delta p_{0}$. The
jump in pressure at the critical point -- when reattachment is at
the trailing edge -- has the following physical explanation. It is
known that the lift drops when the bubble opens, which effectively
means that the pressure balance between the lower and upper
surfaces of an airfoil has changed: some amount of pressure at the
lower surface has leaked into the upper surface, namely into the
bubble. The latter is allowed by unsteadiness of the process,
i.e., the unsteady Kutta-Joukowsky condition.

Therefore, the mechanical analog of a separation bubble is $p =
p_{0} + {\widetilde{\sigma} / R}, \ p
> p_{0}$, so that the bubble grows when the ambient pressure dictated by Bernoulli's equation, $p =
p_{\infty} - {\rho u_{\mathrm{max}}^{2} / 2}$, decreases:
\begin{subequations}
\label{model_mechanistic_hysteresis}
\begin{align}
u_{\mathrm{cr}}^{2}\left( \dot{u}_{\mathrm{max}}
>0 \right) &: \ R_{0} =
\frac{\widetilde{\sigma}}
{
p_{\infty} - p_{0} -
   \frac{
   \rho (u_{\mathrm{cr}}^{2})^{2}
   }
   {2}
}, \\
u_{\mathrm{cr}}^{1}\left(\dot{u}_{\mathrm{max}} <0\right)&: \
R_{0} = \frac{\widetilde{\sigma} }{ p_{\infty} - p_{0} - \Delta
p_{0} - \frac {\rho (u_{\mathrm{cr}}^{1})^{2}}{2}},
\end{align}
\end{subequations}
which produces a hysteretic behavior. Obviously,
$u_{\mathrm{cr}}^{1} < u_{\mathrm{cr}}^{2}$ is consistent with the
physical observations.
\begin{figure}[h!] \centering
\includegraphics[scale=1.0,angle=0]{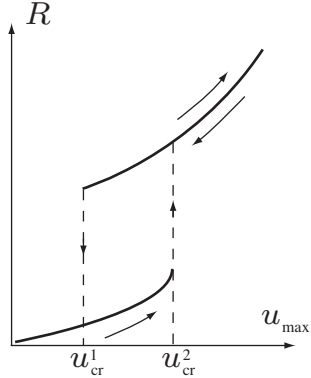}
\caption{\footnotesize Schematics of model \eqref{model_mechanistic_hysteresis}
for hysteresis.}
\label{separation_hysteresis}
\end{figure}

In the light of the above, one can account for the hysteresis in
Figure \ref{conjecture} in model \eqref{simplest_model} as
follows. When $\dot{w}<0$ and $w$ passes through $w_{c}$ the
transformation $w \rightarrow w - \Delta w$ with $\Delta w > 0$ is
applied, since physically the effectiveness of control drops by
$\Delta w$. When $\dot{w}>0$ and $w=w_{c}^{\prime}$ then $w
\rightarrow w + \Delta w$, since too conservative amount of
control has been applied before reaching $w=w_{c}^{\prime}$. These
altogether lead to the desired hysteretic behavior. Same can be
done to account for the hysteretic dependence on the angle of
attack $\alpha$.

The above mechanistic model of the hysteresis captures the physics
and proves that the separation bubble has a nontrivial potential
function associated with it. It should be noted that such type of
discontinuous modeling of hysteresis based on the rate
$\mathrm{sign} \, \dot{w}$ is still widely used in applications
and known as \textit{play} and \textit{stop} (classical Prandtl
model) models, cf. Visintin \cite{Visintin}.

\subsection{\label{subsec:hysteresis_model}Accounting for hysteresis in model \eqref{simplest_model}}

However, for the purpose of deriving a universal model which
combines both the bifurcation and the hysteresis in a dynamic
manner, i.e. suitable for control purposes, it makes sense to
follow another way of modeling hysteresis phenomena, based on the
choice of an appropriate potential function $V(x)$, similar to
what was done in \S \ref{subsec:bubble_bifurcation_model}. We will
enforce this point of view in \S
\ref{subsec:analogies_hysteresis}, where we will illustrate the
analogy to other physical phenomena.

\begin{figure}[h!] \centering
\includegraphics[scale=1.4,angle=0]{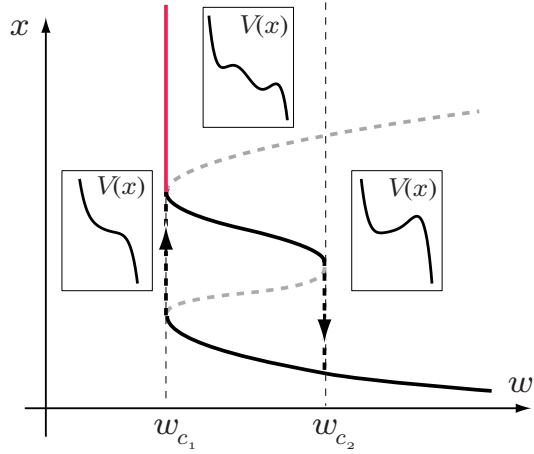}
\caption{\footnotesize Hysteresis
curve in the $(x,w)$-plane and corresponding potential functions;
solid black lines represent stable equilibria, while dashed lines
are unstable equilibria; solid red line represents a dynamic
bifurcation (bubble size grows with time unboundedly). Plots
$V(x)$ in rectangles show the shape of the potential for $w <
w_{c_{1}}$, $w \in [w_{c_{1}},w_{c_{2}}]$, $w > w_{c_{2}}$,
respectively.}\label{hysteresis}
\end{figure}

The grounding thesis is that the true curve of states in Figure
\ref{conjecture} is not the solid discontinuous one, but rather
the ``true'' picture for separation bubbles corresponds to the
smooth curve (including the dashed line) in Figure
\ref{conjecture}, the fact which has not been realized in the
literature before. This smooth curve corresponds to the equilibria
states of an appropriate potential function $V(x)$; the dashed
curve is not physically observable in view of instability of the
corresponding equilibria states. From \S
\ref{subsec:bubble_bifurcation_model} we know that the potential
function should be of special shape, i.e. when $x \rightarrow \pm
\infty$, then $V(x) \rightarrow \mp \infty$, that is the highest
order terms in $\alpha$ should be odd. Then, as a natural
generalization of the picture in Figure \ref{bifurcation}, we
arrive at Figure \ref{hysteresis}.

At the technical level, the lowest order potential suitable for
achieving the picture in Figure \ref{hysteresis} is of the fifth
order, so that model \eqref{simplest_model} becomes
\begin{align}
\ddot{x}(t) = -V_{x}(x;\ldots),
\end{align}
with $V(x)$ of the fifth order. The existence of such potential is
apparent, and its coefficients in the polynomial representation
can be found with the help of linear programming given a set of
inequalities and equalities based on the calibration requirements.

\subsection{\label{subsec:analogies_hysteresis}Analogy to other physical phenomena}

The fact that the hysteresis originates from the particularity of
the potential function is well-known from other physical systems,
e.g. a ferromagnetic drop deforming in a magnetic field
\cite{Bacri:I} and cavitating hydrofoils \cite{Sychev}.

\begin{figure}[ht] \centering
\includegraphics[scale=1.0,angle=0]{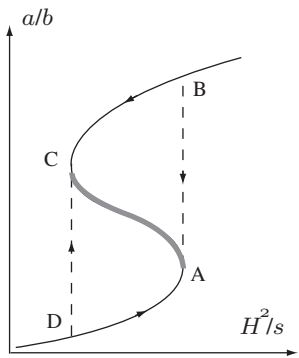}
\caption{\footnotesize Deformation of a ferromagnetic drop.}\label{Ferro}
\end{figure}

Consider the \textit{deformation of ferrofluid drop} of
permeability $\mu_{2}$, placed in a fluid of permeability
$\mu_{1}$, in a magnetic field \cite{Bacri:I,Bacri:II}. The
surface energy of the drop is given by
\begin{align}
E_{s} = \sigma 2 \pi a^{2} e \left[e + \epsilon^{-1}
\sin^{-1}\epsilon\right], \ \epsilon = \left(1-e^{2}\right)^{1/2},
\end{align}
where $e=b/a$ is the aspect ratio, $a$ and $b$ are semi major and
semi minor axes respectively, and $\sigma$ is the interfacial
tension. The magnetic energy is of the form
\begin{align}
E_{m} = - \frac{ V H^{2} }{ 8 \pi}
\frac{\mu_{1} }{ \alpha + n}, \
\alpha = \frac{\mu_{1} }{ \mu_{2} - \mu_{1}},
\end{align}
where $n = e^{2} \left\{- 2 \epsilon +
\log\left[(1+\epsilon)/(1-\epsilon)\right]\right\}/2 \epsilon^{3}$
is the demagnetization factor, $V$ is the volume of the drop, and
$H$ is the applied magnetic field. Minimization of the total
energy, $E_{t}=E_{s}+E_{m}$, with respect to the aspect ratio $e$
produces
\begin{align}
H^{2} / \sigma = g(e),
\end{align}
the behavior of which is depicted in Figure \ref{Ferro}. The
bubble shape is a simple counterplay between magnetic and
interfacial energy of the drop: the former tends to elongate the
drop, while the latter tends to make the drop spherical.
For certain values of $H^{2}/\sigma$ there are three solutions,
but not all of them are stable. When the solution reaches point A
it jumps to the point B, and similarly for the points C and D. AC
portion corresponds to a maximum of $E_{t}$ and thus is unstable,
while the rest of the curve is minima of $E_{t}$ and thus is
stable.

Finally, a \textit{cavitation bubble} on a \textit{hydrofoil},
where hysteresis can be explained with the help of inviscid
free-streamline theory \cite{Acosta:I,Tulin:I}, is another
example, where it has been done analytically. In this physical
problem the boundary of the bubble is well-defined physically and
thus the problem is reliably treated with a free-streamline
theory, that is its predictions \cite{Geurst} agree well with
experiments \cite{Meijer}. For the theoretical treatments of
cavitation flows with free-streamline theory we refer to Tulin
\cite{Tulin:II}, Yeung \& Parkinson \cite{Yeung}, Birkhoff \&
Zarantonello \cite{Birkhoff}, and on the physics of cavitation
flows to Wu \cite{Wu}, Brennen \cite{Brennen}.

\begin{figure}[h!] \centering
\includegraphics[scale=0.90,angle=0]{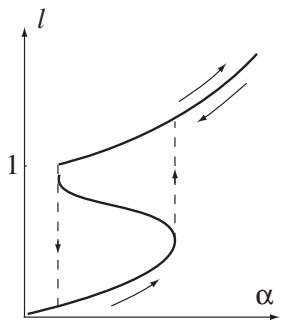}
\caption{\footnotesize Cavitating hydrofoil.}
\label{cavitation_hysteresis}
\end{figure}
On the physical side of cavitation phenomena, it is known from the
general equations of fluid dynamics that the pressure depends on
the velocity distribution (in the steady case) and on the
acceleration (in the unsteady case). More importantly, the
pressure might become negative at points where the velocity is
large. In the majority of cases, fluids cannot sustain a negative
pressure and the continuity of the flow breaks down. As a result,
a region filled with fluid vapor is formed---this is a cavitation
phenomena (see, for instance, \cite{Sedov}). In continuous
incompressible flows the maximum velocity occurs at the
boundary\footnote{This follows from the maximum-modulus theorem
\cite{Titchmarsh}, which states that maxima of a harmonic function
must occur on the boundary, but not in the interior of the
region.} and hence cavitation first appears on the body surface:
\begin{align}
\chi = \frac{2 (p_{st} - p_{d}) }{ 2 u_{\infty}^{2}} =
\frac{u_{\mathrm{max}}^{2} }{ u_{\infty}^{2}} - 1,
\end{align}
which is a cavitation number. Deviations from this law are due to
vortex shedding and other unsteady effects. The behavior of the
cavitation bubble is given by for partially cavitating, $l<1$
\cite{Acosta:I}, and supercavitating, $l>1$ \cite{Tulin:I}, foils
respectively,
\begin{subequations}
\label{cavity_length}
\begin{align}
\frac{\chi }{ 2 \alpha} &= \frac{2 - l + 2 (1-l)^{1/2} }{ l^{1/2}
(1-l)^{1/2}}, \ l < 1, \\
\alpha \left(\frac{2 }{ \chi} + 1\right) &= (1-l)^{1/2}, \ l > 1,
\end{align}
\end{subequations}
where $\alpha$ is the angle of attack. The expressions
\eqref{cavity_length} are basically the solution of the
equilibrium condition, $V' = 0$, and schematically shown in Figure
\ref{cavitation_hysteresis} for fixed cavitation number (see also
Sychev \cite{Sychev}).

\section{Conclusions}

This work has focused on the fundamental aspects---the most
important physics and dynamic behavior---of a generic separation
bubble using thick airfoils as a paradigm. Given an incomplete
experimental knowledge of the complex phenomena of separation
bubble, we applied the deduction based on bifurcation and
catastrophe theory and thus (1) filled in incomplete pieces in the
dynamical picture of the phenomena, (2) advocated that this
dynamical picture is finite-dimensional at the coarse level, (3)
developed a constructive way of building a model, and (4) produced
a model.

The model can be enhanced in particular by (a) incorporating a
non-trivial state equation of a bubble, (b) accounting for
separation at non-zero angle of attack $a_{c} \neq 0$, and (c)
calibrating the model for a given airfoil. These are the future
directions of this study and will require considerable theoretical
and experimental efforts. We also expect that this approach to low
dimensional modeling will be helpful in real time flow control.

\begin{ack}
R.K. would like to thank Prof. Anatol Roshko for helpful and
encouraging discussions. This work was supported in part by NSF-ITR Grant ACI-0204932.
\end{ack}

\bibliographystyle{unsrt}

\begin{thebibliography}{10}

\bibitem{Kailath}
T.~Kailath.
\newblock {\em Linear systems}.
\newblock Prentice-Hall, 1980.

\bibitem{Guckenheimer}
J.~Guckenheimer and P.~Holmes.
\newblock {\em Nonlinear Oscillations, Dynamical Systems, and Bifurcations of
  Vector Fields}.
\newblock Springer, 1983.

\bibitem{Shaw}
R.~S. Shaw.
\newblock {\em The dripping faucet as a model chaotic system}.
\newblock Arial, Santa Cruz, 1984.

\bibitem{Kang}
I.~S. Kang and L.~G. Leal.
\newblock Bubble dynamics in time-periodic straining flows.
\newblock {\em J. Fluid Mech.}, 218:41--69, 1990.

\bibitem{Magill}
J.~Magill, M.~Bachmann, G.~Rixon, and K.~McManus.
\newblock Dynamic stall control using a model-based observer.
\newblock {\em J. Aircraft}, 40:355--362, 2003.

\bibitem{Petot}
D.~Petot.
\newblock Differential equation modeling of dynamic stall.
\newblock {\em Rech. Aerosp.}, 5:59--72, 1989.

\bibitem{Tobak:I}
M.~Tobak, G.~T. Chapman, and L.~B. Schiff.
\newblock Mathematical modelling of the aerodynamic characteristics in flight
  dynamics.
\newblock Technical Report 85880, NASA TM, 1984.

\bibitem{Schlichting}
H.~Schlichting and K.~Gersten.
\newblock {\em Boundary layer theory}.
\newblock Springer-Verlag, 2000.

\bibitem{Jones}
B.~M. Jones.
\newblock Stalling.
\newblock {\em J. Roy. Aero. Soc.}, 38:753--770, 1934.

\bibitem{Pauley}
L.~L. Pauley, P.~Moin, and W.~C. Reynolds.
\newblock The structure of two-dimensional separation.
\newblock {\em J. Fluid Mech.}, 220:397--411, 1990.

\bibitem{Arnold:I}
V.~I. Arnold.
\newblock {\em Bifurcation Theory and Catastrophe Theory}.
\newblock Springer-Verlag, New York, 1999.

\bibitem{Tani}
I.~Tani.
\newblock Low-speed flows involving bubble separations.
\newblock {\em Prog. Aero. Sci.}, 5:70--103, 1964.

\bibitem{Ghil}
M.~Ghil, J.-G. Liu, C.~Wang, and S.~Wang.
\newblock Boundary-layer separation and adverse pressure gradient for 2d
  viscous incompressible flow.
\newblock {\em Physica D}, 197:149--173, 2004.

\bibitem{Oster}
D.~Oster and I.~Wygnanski.
\newblock The forced mixing layer between parallel streams.
\newblock {\em J. Fluid Mech.}, 123:91--130, 1982.

\bibitem{Amitay}
M.~Amitay and A.~Glezer.
\newblock Role of actuation frequency in controlled flow. {R}eattachment over a
  stalled airfoil.
\newblock {\em AIAA J.}, 40:209--216, 2002.

\bibitem{Marsden:I}
J.~E. Marsden and A.~Weinstein.
\newblock {\em Calculus}.
\newblock Springer-Verlag, 1985.

\bibitem{Krechetnikov}
R.~Krechetnikov and I.~I. Lipatov.
\newblock Time-periodic boundary layer under conditions of the large amplitude
  external disturbances.
\newblock {\em Transactions of Central Aero-Hydrodynamics Institute},
  31:27--40, 2000.

\bibitem{Nayfeh}
A.~H. Nayfeh.
\newblock Nonlinear stability of boundary layers.
\newblock In {\em Aerospace Sciences Meeting, 25th, Reno, NV, Jan 12-15}, pages
  1--54, 1987.

\bibitem{Taylor:II}
G.~I. Taylor.
\newblock The formation of emulsions in definable fields of flow.
\newblock {\em Proc. Roy. Soc. London A}, 146:501--523, 1934.

\bibitem{Nishri}
B.~Nishri and I.~Wygnanski.
\newblock Effects of periodic excitation on turbulent flow. separation from a
  flap.
\newblock {\em AIAA J.}, 36:547--556, 1998.

\bibitem{Greenblatt:I}
D.~Greenblatt, B.~Nishri, A.~Darabi, and I.~Wygnanski.
\newblock Dynamic stall control by periodic excitation. {P}art 2: mechanisms.
\newblock {\em J. Aircraft}, 38:439--447, 2001.

\bibitem{Visintin}
A.~Visintin.
\newblock {\em Differential models of hysteresis}.
\newblock Springer, 1994.

\bibitem{Bacri:I}
J.-C. Bacri and D.~Salin.
\newblock Instability of ferrofluid magnetic drops under magnetic field.
\newblock {\em J. Phys. Lett.}, 43:649--654, 1982.

\bibitem{Sychev}
V.~V. Sychev.
\newblock High {R}eynolds number flow past a plate mounted at a small angle of
  attack.
\newblock {\em Fluid Dynamics}, 36:244--261, 2001.

\bibitem{Bacri:II}
J.-C. Bacri and D.~Salin.
\newblock Dynamics of the shape transition of magnetic ferrofluid drop.
\newblock {\em J. Phys. Lett.}, 44:415--420, 1983.

\bibitem{Acosta:I}
A.~J. Acosta.
\newblock A note on partial cavitation of flat plate.
\newblock Technical Report E-19.9, Hydrodynamic Laboratory, California
  Institute of Technology, 1955.

\bibitem{Tulin:I}
M.~P. Tulin.
\newblock Steady two-dimensional cavity flows about slender bodies.
\newblock Technical Report 834, David Taylor Model Basin, 1953.

\bibitem{Geurst}
J.~A. Geurst.
\newblock Linearized theory for partially cavitated hydrofoils.
\newblock {\em Int. Shipbuilding Progr.}, 6:369--384, 1959.

\bibitem{Meijer}
M.~C. Meijer.
\newblock Some experiments on partly cavitating hydrofoils.
\newblock {\em Int. Shipbuilding Progr.}, 6:361--368, 1959.

\bibitem{Tulin:II}
M.~P. Tulin.
\newblock Supercavitating flows -- small perturbation theory.
\newblock {\em J. Ship Res.}, pages 16--37, 1964.

\bibitem{Yeung}
W.~W.~H. Yeung and G.~V. Parkinson.
\newblock On the steady separated flow around an inclined flat plate.
\newblock {\em J. Fluid Mech.}, 333:403--413, 1997.

\bibitem{Birkhoff}
G.~Birkhoff and E.~H. Zarantonello.
\newblock {\em Jets, wakes, and cavities}.
\newblock Academic Press Inc., 1957.

\bibitem{Wu}
T.~Y. Wu.
\newblock Cavity and wake flows.
\newblock {\em Ann. Rev. Fluid Mech.}, 4:243--284, 1972.

\bibitem{Brennen}
C.~E. Brennen.
\newblock {\em Cavitation and bubble dynamics}.
\newblock Oxford Univ. Press, 1995.

\bibitem{Sedov}
L.~I. Sedov.
\newblock {\em Two-dimensional problems in hydrodynamics and aerodynamics}.
\newblock Interscience publishers, 1965.

\bibitem{Titchmarsh}
E.~C. Titchmarsh.
\newblock {\em The theory of functions}.
\newblock Oxford Univ. Press, 1947.

\end{thebibliography}

\end{document}